\documentclass[
    prl, twocolumn, superscriptaddress, nofootinbib, amsmath, amssymb,
    aps
]{revtex4-2}

\usepackage{graphicx,epsfig,psfrag,bm,amssymb,hyperref}
\usepackage{dcolumn}
\usepackage{bm}
\usepackage{color}
\usepackage{soul}
\usepackage{cancel}
\usepackage{mathrsfs,amsfonts,color}
\usepackage[utf8]{inputenc}

\DeclareRobustCommand{\Eq}[1]{Eq.~(\ref{#1})}
\DeclareRobustCommand{\Eqs}[2]{Eqs.~(\ref{#1}) and (\ref{#2})}

\newcommand{\be}{\begin{equation}}
\newcommand{\ee}{\end{equation}}

\newcommand{\VHPD}{V_{\rm HPD}}
\newcommand{\g}{g_{a NN}}
\newcommand{\He}{$^3$He}
\newcommand{\eV}{{\rm eV}}
\newcommand{\rhoDM}{\rho_{{\rm DM}}}
\newcommand{\F}{\mathbb{F}}

\begin{document}

\title{Axion wind detection with the homogeneous precession domain of superfluid helium-3}

\author{Christina Gao}
\affiliation{Department of Physics, University of Illinois Urbana-Champaign, Urbana, IL 61801, USA}
\affiliation{Illinois Center for Advanced Studies of the Universe, University of Illinois Urbana-Champaign, Urbana, IL 61801, USA}
\affiliation{Theoretical Physics Division, Fermi National Accelerator Laboratory, Batavia, IL 60510, USA}
\author{William Halperin}
\affiliation{Department of Physics and Astronomy, Northwestern University, Evanston, IL 60208, USA}
\author{Yonatan Kahn}
\affiliation{Department of Physics, University of Illinois Urbana-Champaign, Urbana, IL 61801, USA}
\affiliation{Illinois Center for Advanced Studies of the Universe, University of Illinois Urbana-Champaign, Urbana, IL 61801, USA}
\author{Man Nguyen}
\affiliation{Department of Physics and Astronomy, Northwestern University, Evanston, IL 60208, USA}
\author{Jan Sch\"{u}tte-Engel}
\affiliation{Department of Physics, University of Illinois Urbana-Champaign, Urbana, IL 61801, USA}
\affiliation{Illinois Center for Advanced Studies of the Universe, University of Illinois Urbana-Champaign, Urbana, IL 61801, USA}
\author{John William Scott}
\affiliation{Department of Physics and Astronomy, Northwestern University, Evanston, IL 60208, USA}
\date\today

\begin{abstract}
Axions and axion-like particles may couple to nuclear spins like a weak oscillating effective magnetic field, the ``axion wind.'' Existing proposals for detecting the axion wind sourced by dark matter exploit analogies to nuclear magnetic resonance (NMR) and aim to detect the small transverse field generated when the axion wind resonantly tips the precessing spins in a polarized sample of material. We describe a new proposal using the homogeneous precession domain (HPD) of superfluid \He\ as the detection medium, where the effect of the axion wind is a small shift in the precession frequency of a large-amplitude NMR signal. We argue that this setup can provide broadband detection of multiple axion masses simultaneously, and has competitive sensitivity to other axion wind experiments such as CASPEr-Wind at masses below $ 10^{-7} ~ \eV$ by exploiting precision frequency metrology in the readout stage.
\end{abstract}

\maketitle

Axions and axion-like particles (ALPs) (denoted $a$) are CP-odd pseudo-Goldstone bosons whose couplings to matter respect a shift symmetry, and which may also make up the cosmic dark matter (DM) density~\cite{Preskill:1982cy,Abbott:1982af,Dine:1982ah}. The ALP coupling to nuclei $N$, $\g \partial_\mu a \overline{N} \gamma^\mu \gamma^5 N$, reduces in the non-relativistic limit to a Hamiltonian $\gamma \vec{B}_a \cdot  \vec{\sigma}_N$, where
\be
\vec{B}_a = \frac{\g}{\gamma}  \nabla a \simeq \g \frac{\sqrt{2 \rhoDM}}{\gamma} \cos(\omega_a t) \vec{v}_a
\ee
acts as an effective oscillating magnetic field which couples to nuclear spins \cite{Graham:2013gfa}. Here $\vec{v}_a \sim 10^{-3}c$ is the DM velocity (in what follows we will set $c = 1$), $\omega_a = m_a (1 + \mathcal{O}(v_a^2))$ with $m_a$ the axion mass, $\rhoDM = 0.3 \ {\rm GeV}/{\rm cm^3}$ is the DM density, and $\gamma$ is the gyromagnetic ratio of $N$. Because $\vec{B}_a$ is proportional to the DM velocity, this interaction is known as the ``axion wind.'' For QCD axions which could resolve the strong CP problem~\cite{Peccei:1977hh,Peccei:1977ur,Weinberg:1977ma,Wilczek:1977pj}, $\g \approx 3 \times 10^{-8} \ {\rm GeV}^{-1}(m_a/{\rm eV})$~\cite{GrillidiCortona:2015jxo}, yielding 
\be
|\vec{B}_a|_{\rm QCD} \simeq 5 \times 10^{-23}\ {\rm T} \left(\frac{m_a}{10^{-7} \ {\rm eV}}\right)
\ee
taking $\gamma = 0.69 \ {\rm GeV}^{-1}$ ($\gamma/(2\pi) = 32.43 \ {\rm MHz}/{\rm T}$)
for the gyromagnetic ratio of \He. Several experiments are aiming to detect the axion wind, including CASPEr-Wind \cite{Garcon:2017ixh,JacksonKimball:2017elr,Aybas:2021nvn} and CASPEr-ZULF~\cite{Wu:2019exd,Garcon_2019}, which use a polarized sample of nuclear spins as the detection medium; comagnetometers using two species of spins to reduce noise \cite{Bloch:2021vnn,Lee:2022vvb}; and QUAX, which uses polarized electron spins \cite{QUAX:2020adt} (see Refs.~\cite{Irastorza:2018dyq,Adams:2022pbo} for a review of other experimental approaches). In a close analogy to nuclear magnetic resonance (NMR), when an external $B$-field is tuned such that the sample Larmor frequency matches $m_a$, the spins will tip away from the external field, yielding a transverse magnetic field proportional to $B_a$ which grows linearly with time up to the axion coherence time, $\tau_a\simeq 2\pi/(m_a v_a^2)$ as long as this is larger than the spin relaxation time.

In this Letter, we describe a new detection mechanism which converts the small amplitude ALP signal into a small \emph{frequency} shift in a large-amplitude NMR signal, allowing the use of precision frequency metrology to evade amplitude noise. The sample is an unusual phase of superfluid \He, the homogeneous precession domain (HPD), which may be understood as a Bose-Einstein condensate (BEC) of spin-1 magnons formed by Cooper pairs of the spin-1/2 helium nuclei in the B-phase of the superfluid~\cite{bunkov2012spin}. We will first review the properties of the HPD (based largely on the review~\cite{bunkov2012spin}), then describe the effect of the axion wind, and finally estimate the sensitivity of the setup to various ALP masses and couplings.

\section{HPD review}

\begin{figure}[t!]
\vspace{-0.15cm}
\includegraphics[width=0.4\textwidth]{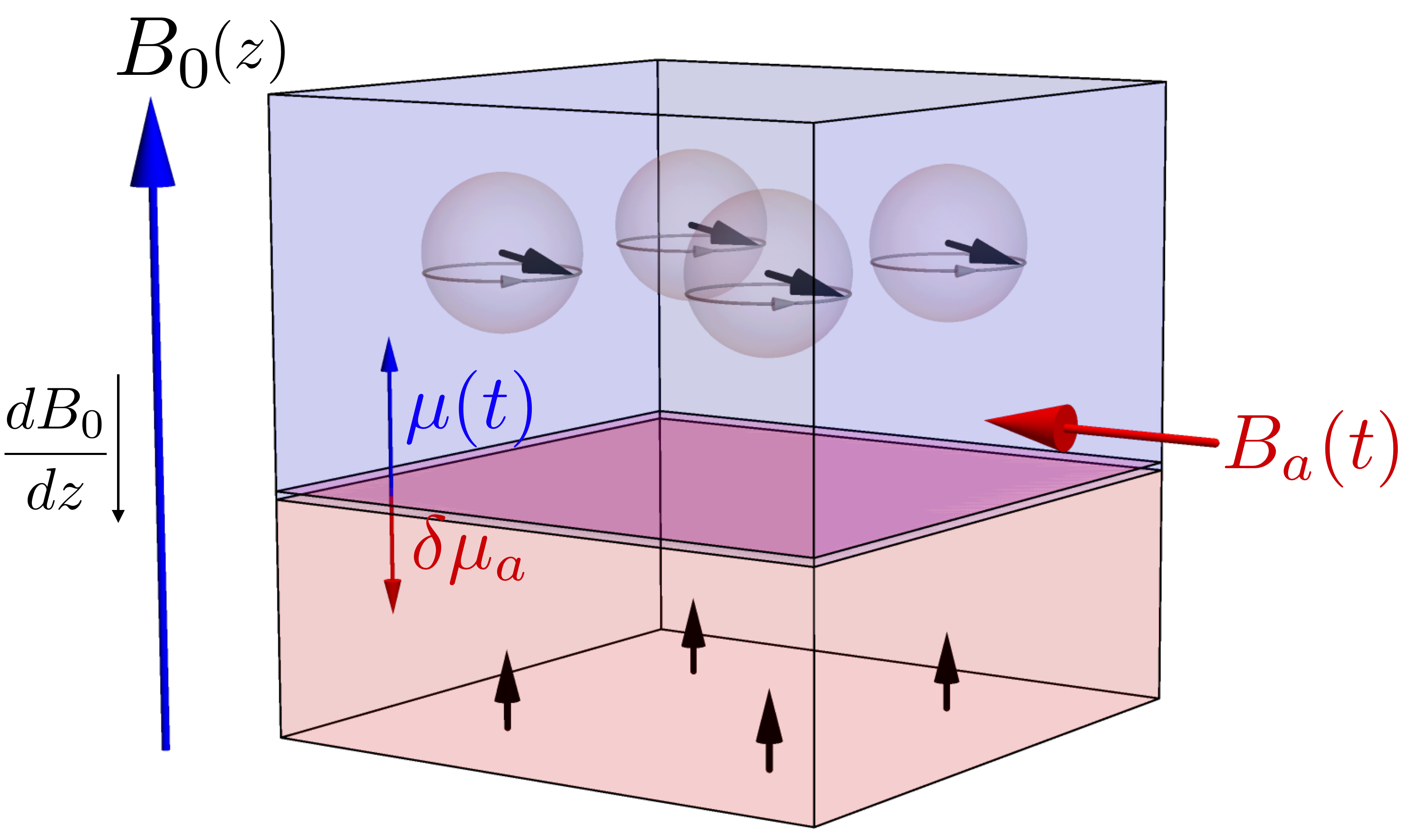}
\caption{ \label{fig:HPD}
Spins in the HPD (blue) precess at the local Larmor frequency set by the location of the DW, while the remainder of the sample (red) is relaxed. A transverse axion wind $B_a$ acts as a small chemical potential $\delta \mu_a$ for magnons.}
\vspace{-0.6cm}
\end{figure}

\He\ is a Fermi liquid which forms a superfluid below temperatures of $\sim$1 K; the excitations of this superfluid are spin-1 magnons. When a static magnetic field $\vec{B}_0 = B_0(z) \hat{z}$ is applied, the equilibrium magnetization is $\vec{M}= \chi \vec{B}_0$, where $\chi \sim 10^{-7}$ is the susceptibility which reaches a maximum at  a temperature of 2.5 mK and pressures of 34 bar \cite{Wheatley_1975}. A transverse magnetic field pulse will torque the spins, injecting some number of spin-1 quanta (i.e.\ magnons), $N_M$. In the B-phase of \He, the HPD is formed by the Bose condensation of these magnons after an RF pulse: a macroscopic fraction of the sample will spontaneously begin to precess coherently about $\vec{B}_0$ with the nuclear spins tipped at the Leggett angle, $\beta_0 = \cos^{-1}(-1/4) \approx 104^\circ$ (originating from dipole-dipole interactions), while the rest of the sample remains relaxed in equilibrium \cite{fomin1984long}. This is illustrated in Fig.~\ref{fig:HPD}: the HPD occupies a volume $\VHPD = N_M/n_M$, where 
\be
n_M = \frac{\chi B_0}{\gamma} (1 - \cos \beta_0) = \frac{5}{4} \frac{\chi B_0}{\gamma}
\ee
is the constant magnon density.

A striking feature of the HPD is that the precession frequency is spontaneously determined. One can view the HPD as a thermodynamic system at fixed particle number $N_M$, where the chemical potential $\mu$ is set by minimizing the free energy $F$. In fact, $\mu$ is the \emph{local} Larmor frequency, corresponding to an angular precession frequency $\omega_{L}(z) = \gamma B_0(z)$, at the location $z$ of the domain wall (DW) separating the HPD from the relaxed domain~\cite{bunkov_bose-einstein_2008}. Thus, even in the presence of a spatially-varying $B_0$, the entire HPD precesses coherently at the \emph{same} frequency. In a lossless system, the precession would continue indefinitely at the frequency established at HPD formation, but due to longitudinal spin relaxation from surface and volume losses on a timescale $T_1$, $N_M$ will decrease monotonically \cite{fomin_separation_1996}. For sufficiently large $T_1$ compared to the spin supercurrent propagation time, the HPD will continuously minimize $F$ at a smaller $\VHPD$, and thus the precession frequency $\omega_L(t)$ will sweep through the local Larmor frequencies corresponding to the motion $z(t)$ of the DW, at a speed governed by $T_1$. $F$ is minimized when the HPD occupies the region of smaller $B$-field, so for $dB_0/dz$ of definite sign, $\omega_{L}(t)$ will decrease monotonically and approximately linearly on timescales much shorter than $T_1$. While the maximum attainable $T_1$ for the HPD is presently unknown, in the low-temperature limit magnetic modes in \He \, have exhibited $T_1$ as long as $1000$ seconds \cite{geller2000stabilization,fisher2000thirty}. The B-phase is destabilized for $B_0 \gtrsim 0.55 \ {\rm T}$~\cite{scholz1981magnetic}, which sets the maximum $\omega_L$ of the HPD.

The aforementioned properties of the HPD may be understood through an effective description of the sample magnetization as described by the Bloch equations. With a static external magnetic field $\vec{B} = B_0(z) \hat{z}$ (not necessarily homogeneous) the magnetization $\vec{M}$ in the HPD evolves as
\begin{align}
\frac{dM_z}{dt} & = \frac{i \gamma}{2} \left( M_{xy} \overline{B}_{xy} - \overline{M}_{xy} B_{xy}\right) - \frac{M_z - \widetilde{M}_0}{5T_1}, \label{eq:BlochMz}\\
\frac{dM_{xy}}{dt} & = -i \gamma \left(M_{xy} B_z - M_z B_{xy} \right) - \frac{M_{xy}}{T_1},
\label{eq:BlochMxy}
\end{align}
where $\widetilde{M}_0$ is the equilibrium magnetization in the HPD, $M_{xy} \equiv M_x + i M_y$ is the transverse magnetization, $\overline{M}_{xy} = M_x - i M_y$, and likewise for $B_{xy}$ and $\overline{B}_{xy}$. The factor of $5 = \frac{\cos \beta_0 - 1}{\cos \beta_0}$ in $T_1$ has been introduced for later convenience; note that the transverse and longitudinal relaxation times are not independent because these components of the magnetization are locked together by the spin supercurrents at the constant tip angle $\beta_0$~\cite{bunkov1995spin}.

Since the magnetization in the relaxed domain does not evolve, we can treat the equilibrium magnetization as belonging to the HPD only via $\widetilde{M}_0 = \chi B_0 \F$, where $\F$ is the HPD fraction of the sample. We parameterize $M_z = \chi B_0 \F \cos \beta_0$ and $M_{xy} = \chi B_0 \F \sin \beta_0 e^{-i \theta}$, which are time-dependent through the HPD fraction $\F(t)$ and because $B_0$ is evaluated at the position $z(t)$ of the domain wall. Suppose first that there is no transverse magnetic field, $B_{xy} = 0$. For homogeneous $B_0$, Eq.~(\ref{eq:BlochMz}) yields
\be
\label{eq:xsolNoAxion}
\F(t) = \F_0 e^{-t/T_1},
\ee 
so the HPD fraction decays exponentially from its initial value $\F_0$ due to relaxation. Similarly, Eq.~(\ref{eq:BlochMxy}) yields
\be
\label{eq:thetasolNoAxion}
\dot{\theta} = \gamma B_0,
\ee
so that precession occurs at the local Larmor frequency $\omega_L = \gamma B_0$.

Due to the DW motion, if the external magnetic field is inhomogeneous, $B_0$ in Eq.~(\ref{eq:thetasolNoAxion}) will acquire an effective time dependence, and thus so will $\omega_L$. To see this, consider the evolution of the HPD over a short time interval $t \ll T_1$. For a fixed cross-sectional area $A$ of the sample container of volume $V$, the position of the domain wall is $z(t)=-h \F(t)$, where $h = V/A$ is the height of the container and we have set $z = 0$ at the top of the container. 
At $t = 0$ we have $z_0= -h\F_0$. 
For short times $t \ll T_1$, the DW position is
\be
z(t)=-h \F_0e^{-t/T_1} \approx z_0 + v_D t,
\ee
where $v_D \equiv h \F_0/T_1 = |z_0|/T_1$ is the instantaneous DW velocity at the height $z_0$. 
Now, since the HPD precesses according to the magnetic field $B_0(z)$ at the DW, we Taylor-expand
\be
B_0(z) \approx B_0(z_0) + \left . \nabla_z B_0 \right |_{z_0} (z - z_0) \equiv B_0(1 +\alpha v_D t),
\ee
where we have defined $\alpha \equiv  \nabla_z B_0 / B_0|_{z_0}$ as a tuneable parameter of the experiment. For the geometry defined in Fig.~\ref{fig:HPD}, $\alpha < 0$, and $\alpha$ can be taken to be constant in the region $z \approx z_0$. We further require $|\alpha (z - z_0)| \ll 1$ so that $n_M$ is approximately constant. 
Combining \Eqs{eq:xsolNoAxion}{eq:thetasolNoAxion} and expanding to first order in $t/T_1$, the transverse magnetization is
\be
M_{xy}(t \ll T_1) = M_{xy}(0) \left(1 - \frac{t}{T_1}\right)e^{-i \omega_L^0(1 + \alpha v_D t/2)t},
\ee
where $\omega_L^0 = \gamma B_0(z_0)$. The linear downward drift of the precession frequency, $\dot{\omega}_L \equiv \ddot{\theta}(t) = \alpha v_D \omega_L^0 < 0$, is the key feature of the HPD. 

\section{Effect of the axion wind on the HPD}

The effect of the axion wind on the HPD evolution can be directly computed by including $\vec{B}_a$ as a source in the Bloch equations. We will first describe the effect qualitatively and estimate its parametric size, and then confirm this behavior with numerical simulations in the following section.

In CASPEr, Larmor resonance is achieved when $\gamma B_0 = m_a$, and $B_a$ generates a tip angle $\delta \beta(t) = \gamma B_a t$ that grows linearly with time up to ${\rm min}[\tau_a, T_2^*]$ where $T_2^*$ is the dephasing time. The three principal challenges in this setup are (i) detecting the extremely small transverse field generated by $\delta \beta$; (ii) tuning $B_0$ in steps of $\sim 10^{-6}$ to achieve resonance at each possible axion mass, given the axion bandwidth $m_a v_a^2 \sim 10^{-6} m_a$; and (iii) generating a uniform $B_0$ to ensure that nearby spins do not dephase, thus preserving $T_2^* > \tau_a$. 
The HPD setup naturally sidesteps all of these challenges, as follows. Because the HPD is a macroscopic quantum state, the inhomogeneous $B$-field, which generates the linear frequency drift,  preserves the precession phase in a macroscopic volume while scanning Larmor frequencies. This allows the HPD to probe many axion masses with the same field profile $B_0$. In the HPD, the tip angle is fixed, so the torque from a transverse $B_a$ instead resonantly changes the number of magnons in the BEC when the local Larmor frequency is equal to $m_a$, $|\Delta N_a| \simeq \chi B_0 B_a \VHPD \tau_a$. The axion wind can thus be seen as a small chemical potential $\delta \mu_a$ for magnons, as shown in Fig.~\ref{fig:HPD}. Since the magnon density in the BEC remains constant, $\VHPD$ changes with $\Delta N_a$, and hence the DW position shifts by an amount $|\Delta z| \simeq  \gamma B_a |z_0| \tau_a$. This leads to a frequency shift,
\be
\label{eq:freqshiftestimate}
\begin{split}
&\left |\frac{\Delta \omega_a}{\omega_L^0}  \right |\simeq ( \gamma B_a)( \alpha z_0) \tau_a\\
&\approx 3\times 10^{-13}\left(\frac{g_{aNN}}{10^{-10}\ \rm GeV^{-1}} \right)\left(\frac{\alpha z_0}{0.02}\right)\left(\frac{10^{-7}\rm eV}{m_a}\right).
\end{split}
\ee 
Note that since $B_a \propto m_a$ along the QCD line, the QCD axion frequency shift is independent of the mass:
\be
\label{eq:freqshiftqcdaxion}
\left ( \left | \frac{\Delta \omega_a}{\omega_L^0} \right | \right)_{\rm QCD}\approx 10^{-17}\left(\frac{\alpha z_0}{0.02}\right).
\ee

While \Eq{eq:freqshiftqcdaxion} is a very small frequency shift, it is comparable to the best sensitivity currently achieved by microwave atomic clocks \cite{doi:10.1126/science.abb2473}, and could thus be detected in principle with homodyne detection of the NMR signal referenced to a local oscillator. In contrast to CASPEr, the overall amplitude of the NMR signal from the HPD is large, of order $M_{xy}(0) \sim \chi B_0$, with the smallness of the axion signal appearing in the frequency rather than the amplitude. 

The resonance can persist while the DW sweeps through a frequency bandwidth up to the axion bandwidth, giving a resonance timescale
\be\label{eq:Deltat}
t_r \approx \frac{10^{-6} m_a}{|\dot\omega_L|} \sim \frac{10^{-6}}{\alpha z_0}T_1.
\ee
Define $\alpha^*$ such that $t_r = \tau_a$, e.g. 
\be
\alpha^* z_0 \simeq 0.02\left( \frac{T_1}{1000 \ \rm sec}\right)\left(\frac{m_a}{10^{-7}\ \rm eV}\right).
\label{eq:alphastar}
\ee
Thus, depending on the choice of $\alpha$,
\be
\left |\frac{\Delta \omega_a}{\omega_L^0} \right | \simeq ( \gamma B_a)( \alpha z_0) 
\left\{
\begin{array}{cl}
t_r~, & t_r \leq \tau_a \ (\alpha \geq \alpha ^*),\\
\sqrt{t_r \tau_a}~, & t_r \geq \tau_a \ (\alpha \leq \alpha^*)~.
\end{array}\right .
\label{eq:SignalFrequencyShiftTimeScaling}
\ee
The scaling with $\sqrt{t_r}$ is a typical feature of measurements exceeding the coherence time and arises from adding in quadrature the incoherent frequency shifts in each interval $\tau_a$ ~\cite{Budker:2013hfa}. 
The signal is maximized when $\alpha=\alpha^*$, which recovers \Eqs{eq:freqshiftestimate}{eq:freqshiftqcdaxion}, but choosing larger $\alpha$ allows more axion masses to be scanned.

\begin{figure*}[t!]
\vspace{-0.5cm}
    \includegraphics[width=\textwidth]{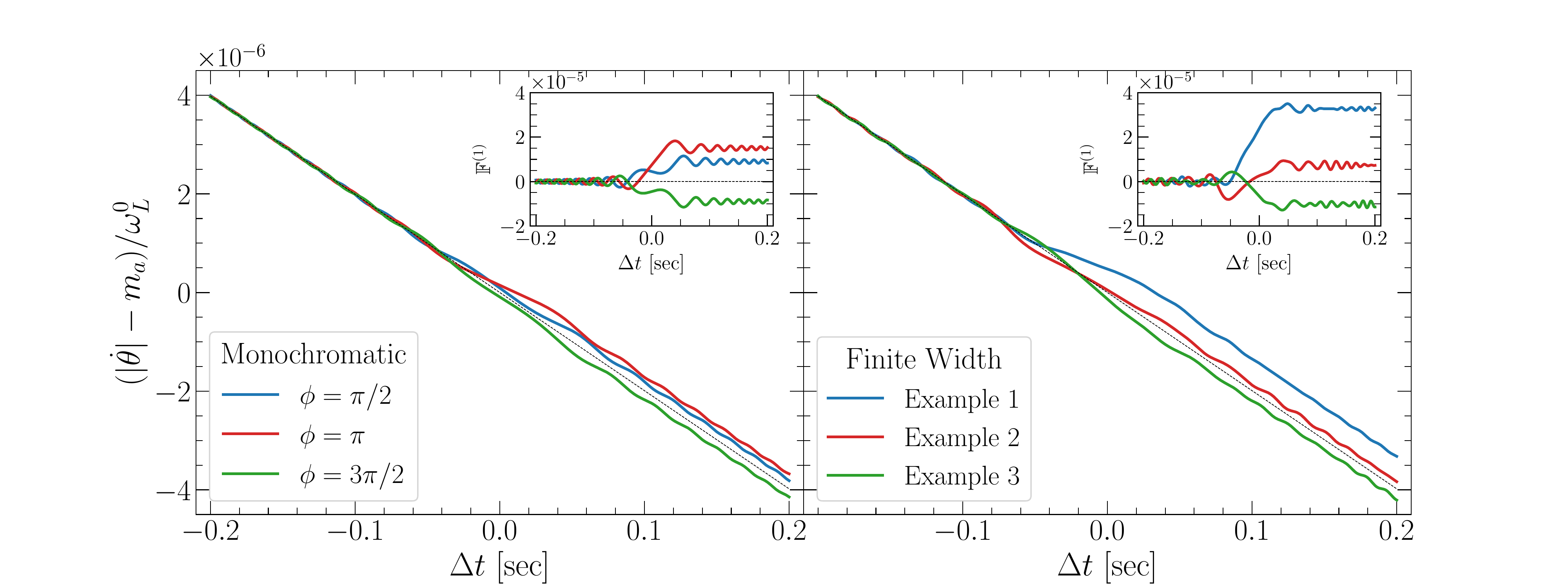}
    \caption{The fractional HPD frequency shift $(|\dot{\theta}| - m_a)/\omega_L^0$ (inset: first-order HPD fraction $\F^{(1)}$) near resonance, with $g_{aNN}=5\times 10^{-5}$ GeV$^{-1}$, $B_0=0.55$ T ($m_a \approx 7\times10^{-8}$ eV), $\alpha z_0\simeq 0.02$, and $T_1=1000$ sec. The axion coupling is taken to be artificially large for illustration purposes to render the frequency shift visible.
   The resonance occurs at $\Delta t = 0$, and $t_r = \tau_a = 0.05 \ {\rm s}$, and $\vec{B}_a$ is assumed to be transverse. The grey dashed lines indicate the behavior in the absence of an axion resonance. \textbf{Left:} monochromatic axion field $\vec{B}_a(t) = \frac{\g \sqrt{2\rhoDM}}{\gamma} \vec{v}_0 \cos(m_at+\phi)$ for three choices of $\phi$. \textbf{Right:} three realizations of the stochastic axion field $\vec{B}_a(t)$ given by Eq.~(\ref{eq:fa}).
 }
\label{fig:x1}
\vspace{-0.2cm}
\end{figure*}

We can confirm the estimate in \Eq{eq:freqshiftestimate} with a perturbative analysis of the Bloch equations. Taking a transverse monochromatic axion field, $B_x = B_a \cos (m_a t + \phi)$, and writing $\F = \F^{(0)} (1+ \F^{(1)})$ where $\F^{(0)}$ is the solution of \Eq{eq:xsolNoAxion} and $\F^{(1)}$ is proportional to $B_a$, we have for $t \ll T_1$ and to leading order in $B_a$,
\be
\label{eq:FirstOrderx}
\frac{d\F^{(1)}}{dt}  \approx  - \tan \beta_0  \gamma B_a \cos (m_a t + \phi) \sin (-\omega_L(t) t).
\ee
On resonance when $m_a = \omega_L(t)$, the right-hand side contains a constant term $\frac{1}{2}\sin \phi$, and thus the solution for $\F^{(1)}$ is
\be
\F^{(1)}(t) \approx -\left(\frac{1}{2} \sin \phi\right) (\gamma B_a \tan \beta_0) t
\ee
which grows linearly with time. In fact, integrating Eq.~(\ref{eq:FirstOrderx}) we can obtain an analytic solution valid for $t \ll T_1$,
\be
\F^{(1)}(t) \approx \left(\frac{-\tan\beta_0\gamma B_a}2\right)
\frac{\cos\phi~ S(\Omega t)+\sin\phi~ C(\Omega t)}{\Omega},
\label{eq:Fresnel}
\ee
where $\Omega\equiv \sqrt{\frac{\alpha z_0 \omega_L^0}{\pi T_1}}$ and $S(x)$ and $C(x)$ are the Fresnel sine and cosine integrals, respectively.

The effect of the axion can be interpreted as a first-order contribution to the DW velocity $\Delta v_D =  -\left(\frac{1}{2} \sin \phi\right)z_0 \tan \beta_0 \gamma B_a$, which then feeds back into the precession frequency because $z - z_0 = h \F_0 - h\F^{(0)}(1 + \F^{(1)}) \approx (v_D + \Delta v_D)t$. In other words, the first-order frequency shift is $\Delta \omega_a(t) = \omega_L^0 \alpha \Delta v_D t$, and integrating up to $\tau_a$ we recover the parametric estimate of \Eq{eq:freqshiftestimate}, up to the $\mathcal{O}(1)$ factor $\frac{1}{2} \sin \phi \tan \beta_0$. There is also an additional first-order effect on the precession phase $\theta$, but it is always parametrically smaller than the frequency shift induced from the variation of $\VHPD$. Finally, note that a longitudinal axion field, $B_z = B_a \cos (m_a t + \phi)$, will not yield a resonance but will rather imprint fast oscillations at frequency $m_a$ on the phase $\theta$.

\section{Simulations, noise, and sensitivity}

To validate the above analysis, we numerically solved the Bloch equations with an axion source, both for a monochromatic axion field and an axion field with the expected $10^{-6}$ bandwidth from the DM velocity dispersion. Since the axion field is a random field with a known power spectrum, for the latter case, we used the time-domain axion field constructed from sampling the speed distribution according to the prescription in Ref.~\cite{Foster:2017hbq}. Specifically, we took
\begin{align}
\vec{B}_a & = \g \frac{ \sqrt{2\rhoDM}}{\gamma} \vec{v}_0 \nonumber \\
& \times  \sum_j\alpha_j\sqrt{f(v_j)\Delta v}\cos\left(m_a\left(1+v_j^2/2\right)t+\phi_j\right).
\label{eq:fa}
\end{align}
The sum is over groups of axion particles, each group having a speed $\in[v_j,v_j+\Delta v]$.  Here $\alpha_j$ is a Rayleigh-distributed random variable, $\phi_j\in[0,2\pi)$ is a random phase, and $f(v)$ is the local DM speed distribution which we take to be Maxwellian with dispersion $v_0 = 220$ km/sec, boosted to the lab frame by $v_{\rm Earth} = 232.24$ km/sec.
A key feature of the HPD setup is the possibility of observing daily modulation, since the same axion masses may be scanned multiple times throughout the day. However, for this analysis we fix $\vec{v}_0$ to be transverse and leave a full daily modulation analysis for future work. Indeed, we use the stochastic field simply to validate our treatment of the coherence time, and we expect that our parametric estimates of the signal strength will differ by $\mathcal{O}(1)$ factors from a full treatment of the correlated components of $\nabla a$ \cite{Lisanti:2021vij,Gramolin:2021mqv}.

Fig.~\ref{fig:x1} shows the evolving fractional frequency shift $(|\dot{\theta}(t)| - m_a)/\omega_L^0$ and the perturbation to the HPD fraction $\F^{(1)}(t)$, when a transverse axion wind is on resonance with the Larmor frequency. The DW can shift either up or down depending on the phase when the resonance occurs, and hence the frequency can shift above or below the linear drift indicated by the grey lines. As mentioned earlier, if the axion wind is longitudinal, no such resonance occurs. Both the monochromatic and the stochastic axion sources show the expected linear growth in $\F^{(1)}$ during the resonance (given by Eq.~(\ref{eq:Fresnel}) in the case of the monochromatic axion). The accumulated frequency shift $\Delta \omega_a/\omega_L^0$ appears as a persistent offset at $\Delta t \gg t_r$ and matches our parametric estimates in Eq.~(\ref{eq:freqshiftestimate}).

To obtain a projected sensitivity for the HPD setup, we need to consider possible noise sources. Since we have already argued that the expected $\Delta \omega_a$ is (in principle) within the sensitivity of state-of-the-art atomic clocks, we will focus on irreducible noise sources. Fortunately, the unique properties of the HPD makes thermal noise a negligible concern. Unlike the case of a BEC of atoms, thermal fluctuations in \He\ do not cause the number of magnons in the condensate to fluctuate, but rather affect the normal fluid component of the \He\ superfluid. Thus, thermal noise will not lead to a frequency shift. 

In fact, the leading irreducible noise source is stochastic fluctuations in magnon number. Over a time interval $\Delta t \ll T_1$, losses reduce the average number of magnons by $N_{\rm loss} = n_M \Delta \VHPD$, where $\Delta \VHPD =  v_D  A\Delta t$ is the change in HPD volume over $\Delta t$. This will lead to Poissonian ``shot noise'' of order $\sigma_{N} = \sqrt{N_{\rm loss}}$, which will cause the HPD volume and hence the frequency to fluctuate.
Taking $\Delta t = \tau_a$, the noise on the frequency shift is
\begin{align}
& \left |\frac{\Delta \omega_{\rm stoch.}}{\omega_L^0}  \right |   = 7 \times 10^{-15} \nonumber \\ 
& \times \left(\frac{\alpha z_0}{0.02}\right) \left(\frac{10^3 \ {\rm s}}{T_1}\right)^{1/2}\left(\frac{10^{-7} \ {\rm eV}}{m_a}\right)\left(\frac{100 \ {\rm cm}^3}{\VHPD}\right)^{1/2}.
\end{align}
We leave the precise microscopic characterization of the stochastic noise to future work, including an analysis of surface and volume loss mechanisms, but we note that the quantized fluctuation of magnons in the HPD as parameterized by $T_1$ is a reasonable coarse-grained characterization of many such microphysical phenomena.

Since the Larmor frequency continuously changes in the experiment, the number of axion resonances covered in a single run is given by roughly $T_1/t_r \sim 10^6(\alpha z_0) $, where $t_r$ is given by \Eq{eq:Deltat}. 
Repeating the experiment $\mathcal{N}$ times suppresses the noise by $\sqrt \mathcal{N}$. However, since the axion field is incoherent between the different runs, the signal-to-noise ratio SNR $\equiv \left |\frac{\Delta \omega_a}{\Delta\omega_{\rm stoch.}}\right |$ scales as $\mathcal{N}^{1/4}$ for the same reasons as discussed in Eq.~(\ref{eq:SignalFrequencyShiftTimeScaling}). Letting $\mathcal{N}=t_{\rm int}/T_1$ where $t_{\rm int}$ is the total integration time, we have
\be
{\rm SNR} \approx \gamma B_a \sqrt{\VHPD n_M}(T_1 t_{\rm int})^{1/4} \times {\rm min}[\sqrt{t_r}, \sqrt{\tau_a}].
\ee
Choosing $t_r = \tau_a$ for the maximum sensitivity gives
\begin{align}
{\rm SNR}_{\rm QCD} & \approx 0.03 \left(\frac{\VHPD}{100 \ {\rm cm}^3}\right)^{1/2} \left(\frac{T_1}{10^3 \ {\rm s}}\right)^{1/4}\left(\frac{t_{\rm int}}{1 \ {\rm yr}}\right)^{1/4} \nonumber \\
& \times \left(\frac{m_a}{7 \times 10^{-8} \ {\rm eV}}\right)
\end{align}
for couplings on the QCD line; the axion mass $m_a = 7 \times 10^{-8} \ {\rm eV}$ is the largest mass which can be probed before the HPD destabilizes at large fields.

\begin{figure}[t]
\vspace{-0.25cm}
    \centering
    \includegraphics[width=0.45\textwidth]{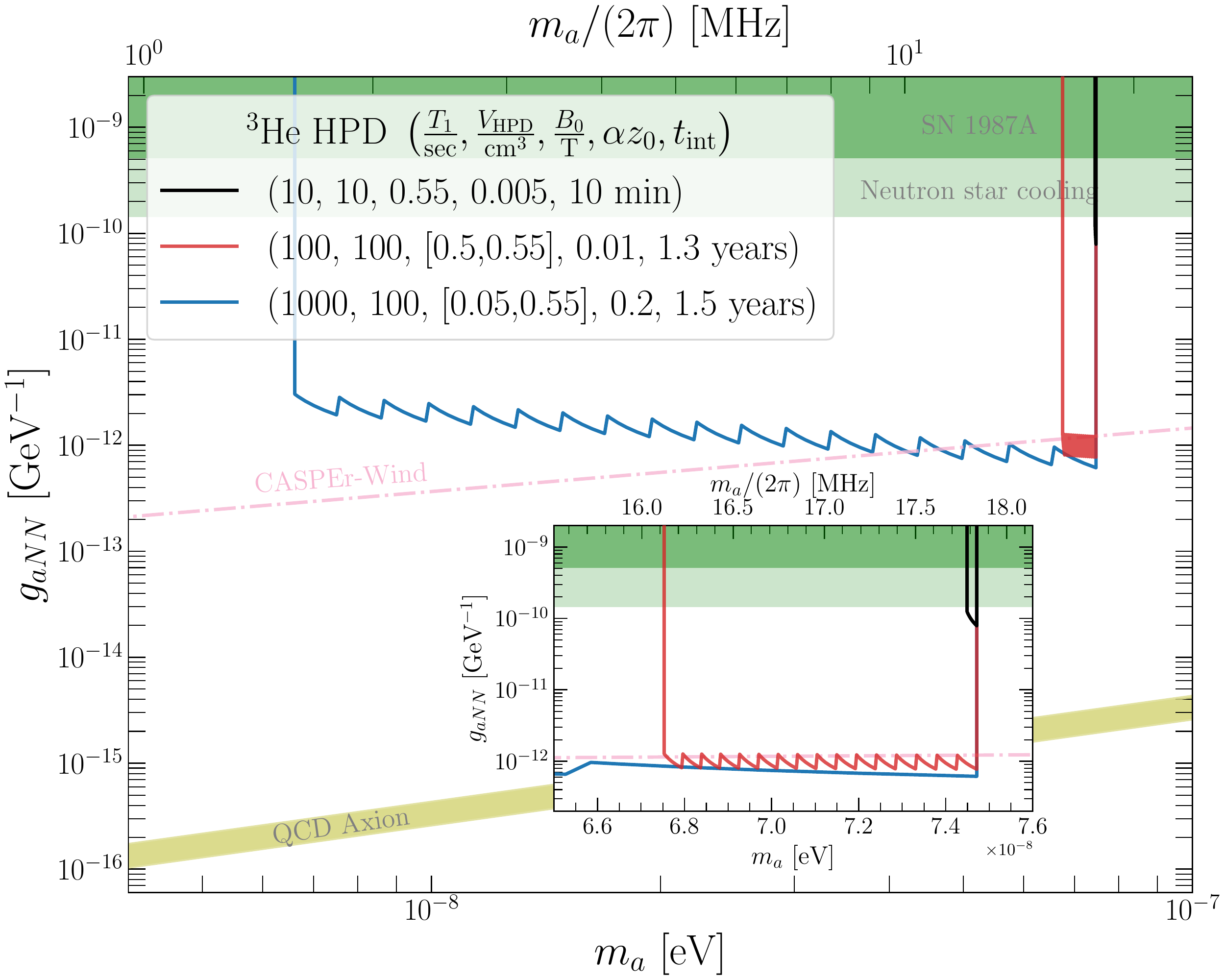}
\caption{
Projected sensitivity to $\g$ (SNR = 1) from the \He\ HPD (black, red, blue), for various choices of parameters. The green shaded regions correspond to the star cooling bounds from SN 1987A~\cite{Carenza_2019}, and neutron stars~\cite{Beznogov_2018} (assuming an axion-neutron coupling). 
The dashed red lines are the CASPEr-Wind projections from Ref.~\cite{JacksonKimball:2017elr}. }
\label{fig:limit}
\vspace{-0.3cm}
\end{figure}

Fig.~\ref{fig:limit} shows a projected sensitivity on $\g$ taking SNR $=1$. Also shown are the star cooling bounds (green) from SN 1987A~\cite{Carenza_2019} and neutron stars~\cite{Beznogov_2018} (assuming an axion-neutron coupling), as well as the projected sensitivity (red) from CASPEr-Wind~\cite{JacksonKimball:2017elr}. As anticipated, choosing a large gradient $\alpha$ results in a larger coverage in axion masses, at the cost of less time spent in each resonance and hence a smaller SNR. However, comparable sensitivity to CASPEr-Wind over an order of magnitude in $m_a$ can be achieved with reasonable parameters in 1.5 years of measurement time at 1 month per $B$-field tune.

\section{Conclusions}

In this Letter we have described a new proposed experiment for detection of the axion wind which exploits the macroscopic coherent properties of the B-phase of \He. In future work we plan to further improve the projected sensitivity by taking into account the correlations between the $\mathcal{N}$ measurements induced by daily modulation, as well as correlations between spatially-separated samples~\cite{Foster:2020fln}, which may increase the SNR such that QCD axion detection at masses $\simeq 10^{-7} \ {\rm eV}$ is possible. We close by emphasizing that even without further improvements to $T_1$ beyond those already achieved in the laboratory (black curve in Fig.~\ref{fig:limit}) the HPD detection scheme could achieve world-leading limits on the axion wind coupling which exceed the most stringent astrophysical bounds.

\emph{Acknowledgments.}  We thank Doug Beck, Andrei Derevianko, Vladimir Eltsov, Jeff Filippini, Joshua Foster, Elizabeth Goldschmidt, Tony Leggett, David J.E. Marsh, Nicholas Rodd, Benjamin Safdi, James Sauls, Ilya Sochnikov, Tomer Volansky, and Kathryn Zurek for helpful discussions. We thank Rachel Nguyen for assistance with implementation of the time-domain axion field, and Yiming Zhong for artistic assistance. We especially thank Andrew Geraci for facilitating collaboration which led to many of the ideas in this work. The work of YK and JSE is supported in part by DOE grant DE-SC0015655. CG acknowledges the Aspen Center for Physics for its hospitality where part of this work is done, which is supported by National Science Foundation grant PHY-1607611. CG is supported by the DOE QuantISED program through the theory consortium “Intersections of QIS and Theoretical Particle Physics” at Fermilab. WPH, MDN, and JWS acknowledge support from the NSF Division of Materials Research grant DMR-2210112.

\vspace{-0.3cm}
\bibliography{HPDBib}

\end{document}